\documentstyle[preprint,eqsecnum,aps]{revtex}
\preprint{UCSBTH-95-27, TIFR/TH/95-46, hep-th/9509108}
\tighten

\newcommand{\be}{\begin{equation}}
\newcommand{\ee}{\end{equation}}
\newcommand{\ben}{\begin{eqnarray}\displaystyle}
\newcommand{\een}{\end{eqnarray}}

\newcommand{\wt}{\widetilde}

\newcommand{\refb}[1]{(\ref{#1})}

\newcommand{\p}{\partial}

\newcommand{\ab}{{(j)}}
\newcommand{\bb}{{(k)}}

\newcommand{\sectiono}[1]{\section{#1}\setcounter{equation}{0}}
\newcommand{\foot}[1]{\footnote{#1}}

\title{ROTATING
BLACK HOLES WHICH SATURATE A BOGOMOL'NYI BOUND}

\smallskip

\author{Gary T. Horowitz}
\address{University of California \\
Santa Barbara, CA 93106, USA \\
gary@cosmic.physics.ucsb.edu }

\author{Ashoke Sen}
\address{Tata Institute of Fundamental Research \\
Homi Bhabha Road, Bombay 400005, INDIA \\
sen@theory.tifr.res.in, sen@tifrvax.bitnet}

\author{September 1995}
\begin{document}
\maketitle

\begin{abstract}

We construct and study the electrically charged, rotating black hole solution
in heterotic string theory compactified on a $(10-D)$
dimensional torus.
 This black hole is characterized by its
mass, angular momentum, and a $(36-2D)$ dimensional electric charge vector.
One of the novel features of this solution is that
for $D >5$, its extremal limit saturates  the Bogomol'nyi
bound. This is in contrast with the $D=4$ case
where the rotating black hole solution develops a naked singularity
before the Bogomol'nyi bound is reached. The extremal black holes
can be superposed, and by taking a periodic array in $D>5$,
one obtains effectively four dimensional solutions without naked singularities.

\end{abstract}

\vfill

\eject

\sectiono{Introduction}

Among the infinite tower of states in string theory,
those which saturate a Bogomol'nyi bound
are of particular interest, since they do not receive
quantum corrections to their mass or charges\cite{OLWI,GIHU,KALL}.
During the past few years, there has been considerable discussion
over whether these states
can be identified with
black holes\cite{DURA,SUSS,EXTRE,PEET,MITRA}.
A necessary condition for this to be
the case is clearly that there exist black hole solutions with the same
values of the mass and charges. For compactifications
of the heterotic string down to four dimensions on a torus, it has been
shown that for all spherically symmetric  BPS saturated states
with $N_L \ne 0$,
there are extremal black hole configurations satisfying this
condition\cite{GIMA,GHS,CVEY,GENBL}.\footnote{The case $N_L = 0$
has recently been discussed \cite{BEHRN,LIKAL,CVESIN}.}
This is encouraging, but there are also
nonspherically symmetric states saturating the bound
which one would like to identify with  rotating black holes. Unfortunately,
in four dimensions, the rotating black hole solutions  become extremal
before  the Bogomol'nyi bound is saturated. The solutions which do
saturate the  bound contain naked singularities, making their physical
significance highly questionable.

We will show that in higher dimensions, this problem disappears (at least
for black holes with a single component of angular momentum). This is
directly related to an unusual property of the higher dimensional
Kerr solution. As shown by Myers and Perry\cite{MYPE}, in dimensions
greater than five,  there are rotating  (uncharged)
black hole solutions with any value
of the ratio $a/m$ where $a$ is the angular momentum parameter and $m$ is
the mass. In other words, there is no extremal limit in this case.
(In four dimensions, solutions with $a>m$ contain
naked singularities.) We will find that when one adds charges to this solution
appropriate to heterotic string theory, there is an extremal limit precisely
when the Bogomol'nyi bound is saturated.

The massless fields in heterotic string theory compactified on a $(10-D)$
dimensional torus consist of the string
metric $G_{\mu\nu}$, the anti-symmetric
tensor field $B_{\mu\nu}$, $(36-2D)$ U(1) gauge fields $A_\mu^\ab$
($1\le j\le 36-2D$), the scalar dilaton field $\Phi$, and a
$(36-2D)\times (36-2D)$
matrix valued scalar field $M$ satisfying,
\be \label{e5}
M L M^T = L, \quad \quad M^T=M.
\ee
Here $L$ is a $(36-2D)\times(36-2D)$
symmetric matrix with $(26-D)$ eigenvalues $-1$ and
$(10-D)$ eigenvalues $+1$. For definiteness we shall take $L$ to be,
\be \label{e4}
L=\pmatrix{-I_{26-D} & \cr & I_{10-D}\cr},
\ee
where $I_n$ denotes an $n\times n$ identity matrix. The action describing
the effective field theory of these massless
bosonic fields is given by\cite{MAHSCH},
\ben \label{e1}
S &=& C\int d^D x \sqrt{-\det G} \, e^{-\Phi} \, \Big[ R_G + G^{\mu\nu}
\p_\mu \Phi \p_\nu\Phi +{1\over 8} G^{\mu\nu} Tr(\p_\mu M L\p_\nu ML)
\nonumber \\
&& -{1\over 12} G^{\mu\mu'} G^{\nu\nu'} G^{\rho\rho'} H_{\mu\nu\rho}
H_{\mu'\nu'\rho'} - G^{\mu\mu'} G^{\nu\nu'} F^\ab_{\mu\nu} \, (LML)_{jk}
\, F^\bb_{\mu'\nu'} \Big] \, ,
\een
where,
\be \label{e2}
F^\ab_{\mu\nu} = \p_\mu A^\ab_\nu - \p_\nu A^\ab_\mu \, ,
\ee
\be \label{e3}
H_{\mu\nu\rho} = \p_\mu B_{\nu\rho} + 2 A_\mu^\ab L_{jk} F^\bb_{\nu\rho}
+\hbox{cyclic permutations of $\mu$, $\nu$, $\rho$}\, ,
\ee
and $R_G$ denotes the scalar curvature associated with the metric
$G_{\mu\nu}$.
$C$ is an arbitrary constant which does not affect the equations of
motion and can be absorbed into the dilaton field $\Phi$.

The general rotating black hole in $D$ dimensions is characterized by
$[(D-1)/2]$ different angular momentum parameters (where $[x]$ denotes the
integer part of $x$). This just corresponds to the
components of angular momentum in different orthogonal two-planes.
We will consider here the simplest case of a single nonzero angular
momentum parameter $a$.  The uncharged rotating black hole in pure
Einstein gravity in $D$ dimensions is given by \cite{MYPE}:
\ben \label{e22}
ds^2 &=& -{\rho^2 + a^2 \cos^2\theta - 2m\rho^{5-D}
\over \rho^2 + a^2 \cos^2 \theta}
dt^2 +{\rho^2 + a^2 \cos^2 \theta \over \rho^2 + a^2 -2m \rho^{5-D}} d\rho^2
+ (\rho^2 + a^2 \cos^2\theta) d\theta^2 \nonumber \\
&& + {\sin^2\theta\over \rho^2 + a^2 \cos^2\theta} [(\rho^2+a^2) ( \rho^2
+ a^2 \cos^2\theta) + 2m\rho^{5-D} a^2\sin^2\theta] d\phi^2 \nonumber \\
&& -{4m\rho^{5-D} a \sin^2\theta \over \rho^2+a^2\cos^2\theta} dt d\phi
+ \rho^2 \cos^2\theta d\Omega^{D-4} \, .
\een
Here $t$, $\rho$, $\theta$, $\phi$ denote four of the space-time coordinates
and
$d\Omega^{D-4}$ is the square of the line element on a $D-4$ dimensional
unit sphere. When $D=4$, this metric reduces to the familiar Kerr  solution.
Notice that, aside from the $d\Omega^{D-4}$ factor, the $D$ dependence always
appears multiplied by the mass in the combination $m \rho^{5-D}$.
The event horizon is located where the $\rho = $ constant surfaces become
null. This implies $G^{\rho \rho}=0$ or
\be \label{e39}
\rho^2 + a^2 -2m\rho^{5-D} =0\, .
\ee
One can immediately see the qualitative difference between $D>5$
and $D\le 5$.
For $D >5$, and $m > 0$, this equation always has a solution
where $\rho$ is positive. Thus an event horizon exists for all values of
$m,a$. Since \refb{e39} has only one (positive) solution in this case,
there is no inner horizon. In contrast, for $D\le 5$, there is a maximal
value of $a$
beyond which the event horizon disappears. The curvature singularity
also changes its character in higher dimensions. For $D>5$, since there
is no inner horizon, the singularity is spacelike. Furthermore, it is easy
to see that the norm of the Killing vector $\partial / \partial t$
diverges at $\rho =0$, showing that this surface is singular. This is
in contrast to the situation in four dimensions where the singularity
is timelike and is concentrated on a ring at $\rho =0$ and $\theta = \pi/2$.
One feature of
this metric which does not change in higher dimensions is that the vector
$\partial / \partial t$ becomes null on a surface outside the horizon,
showing that an ergosphere is present.

Since the metric \refb{e22} is Ricci flat, it is automatically a solution
of the equations of motion derived
from the action \refb{e1} if we also set
\be \label{e6a}
\Phi = 0,  \qquad B_{\mu\nu}=0, \qquad
A^\ab_\mu =0, \qquad M=I_{36-2D}\, .
\ee
In the  next section, we will add a general electric charge to this
solution and discuss its properties. In section 3, we investigate the
extremal limit of this charged, rotating black hole, and show that it
saturates a Bogomol'nyi bound. We also discuss
how one can superpose these higher dimensional
extremal black holes to obtain effectively four dimensional solutions.
Section 4 contains some
concluding remarks.

\sectiono{Rotating, charged black holes in $D\ge 4$}

One can add a general charge to the rotating black hole solution of the
previous section by
applying the solution generating transformations $(O(26-D,1)/O(26-D))
\times (O(10-D,1)/O(10-D))$.\foot{This solution was partially constructed
by Peet\cite{PEET}.}  This generates a nontrivial $\Phi, B_{\mu\nu}$ and
$M$, as well as $A_\mu^\ab$. Since the analysis is
identical to the one given in ref.\cite{GENBL} we shall not give the
details here, but only quote the final result. The solution is given by,
\ben \label{e26}
ds^2 &\equiv& G_{\mu\nu} dx^\mu dx^\nu \nonumber \\
&=&  (\rho^2+a^2\cos^2\theta)\Big\{ - \Delta^{-1}
(\rho^2+a^2\cos^2\theta -2m\rho^{5-D}) dt^2 +
(\rho^2+a^2-2m\rho^{5-D})^{-1} d\rho^2
+d\theta^2 \nonumber \\
&& +\Delta^{-1} \sin^2\theta [\Delta + a^2\sin^2\theta (\rho^2 + a^2\cos^2
\theta + 2m\rho^{5-D} \cosh\alpha\cosh\beta)] \, d\phi^2
\nonumber \\ &&
- 2\Delta^{-1} m\rho^{5-D} a \sin^2\theta (\cosh\alpha + \cosh\beta) dt
d\phi + \rho^2 \cos^2\theta
(\rho^2+a^2\cos^2\theta)^{-1} d \Omega^{D-4} \Big\}\, ,
\een
where,
\be \label{e27}
\Delta = (\rho^2 + a^2\cos^2\theta)^2 + 2m \rho^{5-D}
(\rho^2 + a^2 \cos^2\theta)
(\cosh\alpha \cosh\beta -1) + m^2\rho^{10-2D} (\cosh\alpha -\cosh\beta)^2\, ,
\ee
\be \label{e28}
\Phi={1\over 2} \ln {(\rho^2+a^2\cos^2\theta)^2\over \Delta}\, ,
\ee
\ben \label{e29}
A^\ab_t &=& -{n^\ab\over \sqrt 2} \Delta^{-1} m\rho^{5-D}
\sinh\alpha \{ (\rho^2+
a^2\cos^2\theta)\cosh\beta + m\rho^{5-D} (\cosh\alpha - \cosh\beta)\}
\nonumber \\
&& \quad
\hbox{for} \, \, 1\le j\le 26-D\, , \nonumber \\
&& \nonumber \\
&=& -{p^{(j-26+D)}\over \sqrt 2} \Delta^{-1} m\rho^{5-D}
\sinh\beta \{ (\rho^2+
a^2\cos^2\theta)\cosh\alpha + m\rho^{5-D}
(\cosh\beta - \cosh\alpha)\} \nonumber \\
&& \quad
\hbox{for} \, \, j\ge 27-D\, , \nonumber \\
\een
\ben \label{e30}
A^\ab_\phi &=& {n^\ab\over \sqrt 2} \Delta^{-1} m \rho^{5-D} a \sinh\alpha
\sin^2\theta \{ \rho^2 + a^2\cos^2\theta + m\rho^{5-D}\cosh\beta (\cosh
\alpha -\cosh\beta)\}
\nonumber \\
&& \quad \hbox{for} \, \,  1\le j\le 26-D\, , \nonumber \\
&& \nonumber \\
&=& {p^{(j-26+D)}\over \sqrt 2} \Delta^{-1} m \rho^{5-D} a \sinh\beta
\sin^2\theta \{ \rho^2 + a^2\cos^2\theta + m\rho^{5-D}\cosh\alpha (\cosh
\beta -\cosh\alpha)\} \nonumber \\
&& \quad \hbox{for} \, \,  j\ge 27-D\, , \nonumber \\
\een
\be \label{e31}
B_{t\phi} = \Delta^{-1} m\rho^{5-D} a
\sin^2\theta (\cosh\alpha -\cosh\beta)
\{ \rho^2 + a^2\cos^2\theta + m\rho^{5-D}(\cosh\alpha \cosh\beta -1)\} \,
\nonumber \\
\ee
\be \label{e32}
M = I_{36-2D} +\pmatrix{ P nn^T & Q n p^T \cr Q p n^T & P pp^T\cr}\, ,
\ee
where,
\ben \label{e33}
P &=& 2\Delta^{-1} m^2\rho^{10-2D} \sinh^2\alpha \sinh^2\beta \, ,
\nonumber \\
Q &=& - 2 \Delta^{-1} m\rho^{5-D} \sinh\alpha \sinh\beta \{ \rho^2 + a^2
\cos^2\theta + m\rho^{5-D} (\cosh\alpha \cosh\beta -1) \} \, .
\een
Here $\alpha$ and $\beta$ are two boost angles, $\vec n$ is a $(26-D)$
dimensional unit vector, and $\vec p$ is a $(10-D)$ dimensional unit
vector.

There are several consistency checks on the solution \refb{e26} - \refb{e33}.
First, the solution generating transformation applied to a metric like
\refb{e22} only changes the  $t$ and $\phi$ components of the string metric.
Comparing
\refb{e26} with \refb{e22} we see that indeed the $\rho \rho, \theta \theta$
and additional $D-4$ components of the metric are identical.  Second,
since the $tt, t\phi$ and $\phi\phi$ components of the higher dimensional
Kerr metric depend on $D$ and $m$ only through the combination $m\rho^{5-D}$,
one should be able to obtain the general solution \refb{e26} - \refb{e33}
by starting with the general four dimensional rotating
black hole solution\cite{GENBL} and replacing $m\rho$ with $m\rho^{5-D}$.
This is indeed the case.  Finally, setting the angular momentum parameter
$a$ to zero, one obtains the general electrically charged nonrotating
black hole in $D$ dimensions\cite{PEET}.

{}From eqs.\refb{e26} and \refb{e28} we can also find an expression for
the canonical Einstein metric $g_{\mu\nu}\equiv e^{-2\Phi/(D-2)} G_{\mu\nu}$:
\ben \label{e34}
ds_E^2 &\equiv& g_{\mu\nu} dx^\mu dx^\nu  \nonumber \\
&=& \Delta^{1\over D-2} (\rho^2+a^2\cos^2\theta)^{D-4\over
D-2}\Big\{ - \Delta^{-1}
(\rho^2+a^2\cos^2\theta -2m\rho^{5-D}) dt^2 \nonumber \\
& & + (\rho^2+a^2-2m\rho^{5-D})^{-1} d\rho^2
+d\theta^2 \nonumber \\
&&   +\Delta^{-1} \sin^2\theta [\Delta + a^2\sin^2\theta
(\rho^2 + a^2\cos^2
\theta + 2m\rho^{5-D} \cosh\alpha\cosh\beta)] \, d\phi^2
\nonumber \\  &&
- 2\Delta^{-1} m\rho^{5-D} a \sin^2\theta (\cosh\alpha + \cosh\beta) dt
d\phi + \rho^2\cos^2\theta(\rho^2 + a^2\cos^2\theta)^{-1}
d\Omega^{D-4}\Big\}\, . \nonumber \\
\een

The total mass $M$, angular momentum $J$, and electric charge $Q^\ab$
of these black holes can be obtained from the asymptotic form of the
solution and are given by
\be \label{e35}
M = {1\over 2} m [ 1 + (D-3)\cosh\alpha \cosh \beta]\, ,
\ee
\be \label{e36}
J = {1\over 2} ma (\cosh\alpha + \cosh \beta) \, ,
\ee
\ben \label{e37}
Q^\ab &=&
{m\over \sqrt 2} (D-3) \sinh\alpha \cosh\beta \, n^\ab \qquad \hbox{for}
\quad 1\le j \le 26-D\, \nonumber \\
&=&  {m\over \sqrt 2} (D-3)
\sinh\beta \cosh\alpha \, p^{(j-26+D)} \qquad \hbox{for}
\quad j \ge 27-D\, .
\een
Let us define,
\be \label{elr}
Q^\ab_{L\atop R} = {1\over 2} (I_{36-2D} \mp L)_{jk} Q^\bb\, .
\ee
$\vec Q_L$ and $\vec Q_R$ may be regarded as $(26-D)$ and $(10-D)$
dimensional vectors respectively.  Eq.\refb{e37} gives,
\be \label{eqexp}
(\vec Q_R)^2 = {m^2\over 2}
(D-3)^2
\sinh^2\beta \cosh^2\alpha, \qquad
(\vec Q_L)^2 = {m^2\over 2}
(D-3)^2
\cosh^2\beta \sinh^2\alpha.
\ee

We now consider some properties of these black holes.
Since the $\rho\rho$ component of the string metric is not changed by
the addition of charge,
the event horizon of the solution is again given by \refb{e39}: $\rho_H^2
+a^2 =2m\rho^{5-D}_H  $. (Since the
conformal factor is regular on the horizon, this is also the location of
the event horizon in the Einstein metric.) The area
of the event horizon, i.e. $D-2$ volume,
can be computed from \refb{e34}. This calculation is
simplified by noticing that, on the horizon,  the expression in brackets
in $g_{\phi\phi}$ reduces to
$(\rho_H^2 +a^2)^2(\cosh\alpha
+\cosh\beta)^2/4$.
The area turns out to be
\be\label{area}
A_H = m \rho_H \Omega_{D-2} (\cosh\alpha+\cosh\beta) \ ,
\ee
where $\Omega_{D-2}$ is the volume of a $(D-2)$ sphere of unit radius.
The angular velocity
$\Omega_H$ of the black hole is defined by the condition that $\xi = \p/\p t +
\Omega_H \p/\p \phi$ be null on the horizon. One finds that
\be\label{angvel}
\Omega_H = {a \rho_H^{D-5} \over m (\cosh\alpha+\cosh\beta)} \ .
\ee
We next consider the surface gravity $\kappa$ of the black hole. This can
be obtained from $\xi^\mu \xi_\mu \equiv - \lambda^2$ via
$\kappa^2 = \lim_{\rho \to \rho_H} \nabla_\mu \lambda \nabla^\mu \lambda$.
Since $\kappa$ is a constant on the horizon, one can evaluate it at any point.
It is convenient to choose a point where $\theta = 0$. One obtains
\be\label{sg}
\kappa = {(D-3) \rho_H^2 + (D-5) a^2 \over 2m \rho_H^{6-D}
(\cosh\alpha+\cosh\beta)} \ .
\ee
The Hawking temperature is related to this surface gravity by $T =
\kappa/2\pi$.

Finally, we consider the singularity structure of these black holes for $D>5$.
We first discuss the general case $\alpha \ne \beta$.
Unlike
the vacuum solution, the norm of the Killing vector $\p /\p t$ does not
diverge at $\rho = 0$ but now vanishes there
in both the string and Einstein metrics.
However,  one can see that this surface still contains
a curvature singularity as follows. Eq.
\refb{e28} implies that the dilaton diverges at $\rho = 0$
like $\Phi \approx (D-5) \ln \rho$. In the Einstein metric, this corresponds
to a diverging energy density, which implies a divergence in the Ricci tensor
as $\rho \to 0$. This singularity remains in the string metric as well.
For  $\alpha = \beta$, the Killing vector $\p /\p t$ diverges at
$\rho =0$ in the Einstein metric, showing that this surface is again singular.
In contrast, all components of the string metric now have
regular limits
at $\rho =0, \ \theta \ne \pi/2$, except for the $D-4$ sphere which
shrinks to zero volume (and $G_{\rho\rho}$ which can be made regular by
introducing a new radial coordinate).  However the shrinking spheres
are sufficient to cause the curvature to diverge at $\rho =0$.
Notice that the string coupling
$g = e^{\Phi /2} \rightarrow 0$ near the singularity
as expected for an electrically charged
black hole.

\sectiono{The extremal limit and Bogomol'nyi bound}

Supersymmetry of the toroidally compactified heterotic string theory
implies an upper limit on the ratio of the charge to
the mass known as the Bogomol'nyi bound. It is saturated when
\be \label{ebogo}
M^2 = {1\over 2} \vec Q_R^2\, .
\ee
Eqs.\refb{e35} and \refb{eqexp} show that the only way to satisfy this
equation is to take the limit $m \to 0, \ \beta\to\infty$ keeping
$m_0\equiv m\cosh\beta, \ a$ and $\alpha$ fixed.  For $D \le 5$, there is a
minimum value of $m$ required in order for a solution to \refb{e39}
to exist. Thus
the horizon disappears before the Bogomol'nyi bound is reached. However,
for $D>5$, a horizon exists for all $m>0$.  As we take the limit
$m\to 0$, the location of the horizon, $\rho_H$, approaches the
singularity at $\rho=0$, showing that one cannot increase the charge beyond
this limit.  {\it So for $D>5$,
the solution saturating the Bogomol'nyi bound
is an extreme black hole.}
{}From the above formulas, it is clear that in this
limit, the horizon area and angular velocity go to  zero. The behavior of
the surface
gravity (or Hawking temperature) is rather surprising. If we first set
$a=0$ in \refb{sg} and consider the nonrotating black hole, we see that
the surface gravity approaches a nonzero constant in $D=4$ and vanishes
for $D>4$.
However, for the rotating black hole, we see that the ``critical dimension"
is increased by two: the surface gravity
approaches a nonzero constant for $D=6$ and vanishes for $D>6$.
This is another illustration of the fact that an arbitrarily small amount
of angular momentum can qualitatively change the properties of extreme
dilatonic black holes\cite{HORHOR,ROTAT}.

The solution for $M^2 = \vec Q_R^2/2$ is given by
\ben \label{e26a}
ds^2 &=& (\rho^2+a^2\cos^2\theta) \Big\{ - \Delta^{-1}
(\rho^2+a^2\cos^2\theta) dt^2 +
(\rho^2+a^2)^{-1} d\rho^2
+d\theta^2 \nonumber \\
&& +\Delta^{-1} \sin^2\theta [\Delta + a^2\sin^2\theta (\rho^2 + a^2\cos^2
\theta +2m_0\rho^{5-D}\cosh\alpha)] \, d\phi^2
\nonumber \\  &&
- 2\Delta^{-1} m_0\rho^{5-D} a \sin^2\theta dt
d\phi + \rho^2\cos^2\theta(\rho^2 +
a^2\cos^2\theta)^{-1}d\Omega^{D-4}\Big\}\, .
\een
\be \label{e27a}
\Delta = (\rho^2 + a^2\cos^2\theta)^2 + 2m_0 \rho^{5-D}\cosh\alpha \,
(\rho^2 + a^2 \cos^2\theta)
+ m_0^2\rho^{10-2D} \, ,
\ee
\be \label{e28a}
\Phi={1\over 2} \ln {(\rho^2+a^2\cos^2\theta)^2\over \Delta}\, ,
\ee
\ben \label{e29a}
A^\ab_t &=& -{n^\ab\over \sqrt 2} \Delta^{-1} m_0\rho^{5-D}
\sinh\alpha (\rho^2+ a^2\cos^2\theta)
\quad \hbox{for} \, \, 1\le j\le 26-D\, , \nonumber \\
&=& -{p^{(j-26+D)}\over \sqrt 2} \Delta^{-1} m_0\rho^{5-D} \{ (\rho^2+
a^2\cos^2\theta)\cosh\alpha + m_0\rho^{5-D} \} \quad
\hbox{for} \, \, j\ge 27-D\, , \nonumber \\
\een
\ben \label{e30a}
A^\ab_\phi &=& -{n^\ab\over \sqrt 2} \Delta^{-1} (m_0)^2 \rho^{10-2D}
a \sinh\alpha \sin^2\theta \quad \hbox{for} \, \, 1\le j\le 26-D
\, , \nonumber \\
&=& {p^{(j-26+D)}\over \sqrt 2} \Delta^{-1} m_0 \rho^{5-D} a
\sin^2\theta \{ \rho^2 + a^2\cos^2\theta + m_0\rho^{5-D}\cosh\alpha
\} \quad \hbox{for} \, \,   j\ge 27-D\, , \nonumber \\
\een
\be \label{e31a}
B_{t\phi} = - \Delta^{-1} m_0\rho^{5-D} a
\sin^2\theta
\{ \rho^2 + a^2\cos^2\theta + m_0\rho^{5-D}\cosh\alpha \} \,
\nonumber \\
\ee
\be \label{e32a}
M = I_{36-2D} +\pmatrix{ P nn^T & Q n p^T \cr Q p n^T & P pp^T\cr}\, ,
\ee
\ben \label{e33a}
P &=& 2\Delta^{-1} m_0^2\rho^{10-2D} \sinh^2\alpha \, ,
\nonumber \\
Q &=& - 2 \Delta^{-1} m_0\rho^{5-D} \sinh\alpha \{ \rho^2 + a^2
\cos^2\theta + m_0\rho^{5-D} \cosh\alpha \} \, .
\een
Notice that the ergosphere has disappeared; $\p/\p t$ is now timelike
everywhere.
What is the nature of the singularity at $\rho=0$? A key property is
whether this singularity is timelike or null. Since there is no event
horizon, a timelike singularity would be naked and classical evolution
would break down (although see \cite{HORMAR}). A null singularity is much
more mild.  The criterion for a singularity to be timelike is the
existence of null
geodesics which reach it staying in the past of a $t=$ constant
surface. Consider first the
nonrotating case, obtained by setting $a=0$ in the string metric \refb{e26a}.
Radial null geodesics satisfy $dt = \pm \Delta^{1/2} \rho^{-2} d\rho \approx
\rho^{3-D} d\rho$ near $\rho = 0$. Thus, for all $D\ge 4$, $t$ diverges
along the null geodesic as $\rho \to 0$ showing that the singularity is null.
For the rotating solution with $a\ne 0$, one can consider radial null
geodesics along the rotation axis $\theta = 0$. These satisfy
$dt = \pm \Delta^{1/2}(\rho^2 + a^2)^{-1} d\rho \approx \rho^{5-D} d\rho$
near $\rho = 0$. Thus the singularity is null (at least
in this direction) only for
$D\ge 6$.
These are precisely the dimensions for which
\refb{e26a} describes the extremal limit of a black hole.
The $D=4,5$ solutions contain naked singularities.

Since objects which saturate a Bogomol'nyi bound have no force between
them,
we can also construct stationary
multiple  black hole solutions.
To do this, it is convenient to bring the above solution
into the IWP form\cite{IWP,KALLOSH1}. For simplicity,
we shall consider only the case $\alpha=0$; the $\alpha\ne 0$ solution
can be found by starting from the $\alpha=0$ solution and performing
a boost along one of the internal directions of the $(10-D)$
dimensional torus. We introduce
new Cartesian coordinates $x^1, \ldots x^{D-1}$ through the relations:
\ben \label{edefcor}
x^1 & = & \sqrt{\rho^2+a^2} \sin\theta \cos\phi\, , \nonumber \\
x^2 & = & \sqrt{\rho^2+a^2} \sin\theta \sin\phi\, , \nonumber \\
x^3 & = & \rho \cos\theta \cos \psi^1 \, , \nonumber \\
x^4 & = & \rho \cos\theta \sin \psi^1 \cos \psi^2 \, , \nonumber \\
& \cdot & \nonumber \\
& \cdot & \nonumber \\
x^{D-2} & = & \rho \cos\theta \sin \psi^1 \cdots \sin \psi^{D-5} \cos
\psi^{D-4} \, , \nonumber \\
x^{D-1} & = & \rho \cos\theta \sin \psi^1 \cdots \sin \psi^{D-5} \sin
\psi^{D-4} \, ,
\een
where $\psi^1, \ldots \psi^{D-4}$ are the angles labeling points on
the $(D-4)$ sphere. In this coordinate system the $\alpha=0$ solution
may be written as,
\be \label{enx1}
ds^2 = - F^2(\vec x)) [dt+ \omega_i(\vec x) dx^i]^2
+  d\vec x^2\, ,
\ee
\be \label{enx2}
\Phi =  \ln F(\vec x)\, ,
\ee
\ben \label{enx3}
A^\ab_t & = & 0 \qquad \hbox{for} \qquad 1\le j\le 26-D\, , \nonumber \\
        & = & {p^{(j-26+D)}\over \sqrt 2}
 [F - 1]  \qquad \hbox{for}
\qquad j \ge 27-D \, ,
\een
\ben \label{enx4}
A^\ab_i & = & 0 \qquad \hbox{for} \qquad 1\le j\le 26-D\, , \nonumber \\
        & = & {p^{(j-26+D)}\over \sqrt 2} F
\omega_i \qquad \hbox{for}
\qquad j \ge 27-D \, ,
\een
\be \label{enx5}
B_{ti} = - F \omega_i \, , \qquad M = I_{36-2D}\, ,
\ee
where
\be \label{enx6}
F^{-1}  =  1 +   {  m_0 \rho^{5-D}
\over \rho^2 + a^2 \cos^2 \theta}
\, , \qquad
\omega   =
{m_0 \rho^{5-D} a \sin^2\theta\over \rho^2+a^2\cos^2\theta} d \phi \, .
\ee

The explicit form of  $F$ and $\omega$ in terms of the Cartesian  coordinates
$x^i$ can be obtained  by inverting \refb{edefcor}.
Let $R$ be the radial distance
from the origin in the $D-1$ dimensional Euclidean space, and let $r$
be the radial distance in the $D-3$ dimensional subspace orthogonal to
$x^1$ and $x^2$. Then
\be \label{enx8}
R^2 \equiv \sum_{i=1}^{D-1} (x^i)^2 = \rho^2 + a^2 \sin^2\theta \, ,
 \qquad r^2 =
\sum_{i=3}^{D-1} (x^i)^2 = \rho^2 \cos^2 \theta \, .
\ee
These can be inverted to yield
\ben\label{Rrdef}
 \rho^2 & = &   { (R^2 - a^2) + \sqrt{(R^2 - a^2)^2
 + 4 a^2 r^2} \over 2} \nonumber \\
 a^2 \cos^2 \theta & = & { (a^2 - R^2) + \sqrt{(R^2 - a^2)^2
 + 4 a^2 r^2} \over 2} \, .
\een
Also, from \refb{edefcor}, $\phi=\tan^{-1}(x^2/x^1)$.
Note that $\rho^2 =0$ everywhere on the two dimensional disk $r=0, \ R \le a$,
while $\rho^2+ a^2 \cos^2 \theta = 0$ only on the ring $r=0, \ R = a$.
Thus in the $\vec x$ coordinate system, the singular surface ($\rho=0$)
corresponds to
\be \label{ell2}
x^i=0 \, \, \,\, \hbox{for} \, \, \, \, i\ge 3, \qquad
(x^1)^2+(x^2)^2\le a^2\, .
\ee
In the string metric, this corresponds to a disk of radius $a$.

$F^{-1}$ is a harmonic function
\be \label{enx10}
\sum_{i=1}^{D-1}\p_i \p_i F^{-1} = 0\,
\ee
while $\omega$ satisfies the equation
\be \label{enx7}
 \sum_{i=1}^{D-1}\p_i \p_{[i} \omega_{j]} =0 \ .
\ee
Conversely, any solution of the form given above, where $F^{-1}$ is
an arbitrary harmonic function and $\omega$ is an arbitrary
1-form satisfying eq.\refb{enx7}
will be a solution of the equations of motion. Since both $F^{-1}$ and
$\omega$ satisfy linear equations, it is now easy to construct multi-black
hole solutions by superposing single black hole solutions. If we write
the above single black hole solution as $F^{-1} = 1 + f(\vec x, m_0, a)$,
and $\omega = g(\vec x, m_0, a)$, a general linear superposition is given by
\be \label{enx12}
F^{-1} (\vec x)
=  1 +  \sum_{s=1}^n f(\vec x - \vec x_s, m_s, a_s)\, ,
\ee
and
\be \label{enx13}
\omega(\vec x) = \sum_{s=1}^n g(\vec x - \vec x_s, m_s, a_s)\, .
\ee
$m_s$, $a_s$ and $\vec x_s$ are arbitrary parameters labeling the mass,
angular momentum  and the position of individual black holes.
In this superposition, all of the black holes are spinning in the same
$x^1, x^2$ plane. Also, the electric charge vectors associated with
all the black holes are parallel to each other.

The ten dimensional form of these $D$ dimensional solutions falls into
a class of configurations called `chiral null models' which were introduced
in\cite{HORTSE2} and further studied
in\cite{BEHRN,BEHR}.
They are characterized by a null
translational symmetry and chiral coupling of the worldsheet to the
background. It was shown in\cite{HORTSE2} that these configurations
are exact string solutions and do not receive $\alpha'$ corrections
(in a particular renormalization scheme).

A particularly interesting class of multi-black hole solutions is an
infinite periodic array of black holes.  This can be interpreted as a
solution in a theory where one of the $D-1$ spatial directions has been
compactified\cite{MYERS,KHURI,GAUHAR}\footnote{We wish to
thank J. Schwarz for this suggestion.}.
In order to see how this works, let us consider a periodic array of these
solutions along the direction $x^{D-1}$. Let us define,
\be \label{ekk1}
s^2 = (x^1)^2 + (x^2)^2\, ,
\ee
and
\be \label{ekk2}
\wt R^2 = \sum_{i=1}^{D-2} (x^i)^2\, , \qquad
\wt r^2 = \sum_{i=3}^{D-2} (x^i)^2\, , \qquad
\wt s^2 = s^2 = (x^1)^2 + (x^2)^2\, .
\ee
Now, for a single extremal black hole solution, the asymptotic values of
$F^{-1}$
and $\omega_\phi$ are given by,
\be \label{ekk4}
F^{-1}\simeq 1 +{m_0\over R^{D-3}}\, , \qquad
\omega_\phi \simeq {m_0 a s^2 \over R^{D-1}}\, .
\ee
Thus, for a periodic array of black holes in the $x^{D-1}$ direction
with periodicity one, the asymptotic values of $F^{-1}$ and $\omega_\phi$
are given by,
\be \label{ekk5}
\wt F^{-1} \simeq 1 + \sum_{n=-\infty}^\infty {m_0 \over \{\wt R^2 +
(x^{D-1}-n)^2\}^{D-3\over 2}}\, ,
\ee
\be \label{ekk6}
\wt \omega_\phi \simeq \sum_{n=-\infty}^\infty {m_0 a \wt s^2 \over
\{ \wt R^2 + (x^{D-1}-n)^2\}^{D-1\over 2}}\, .
\ee
For large $\wt R$ the summand is a slowly varying function of $n$ and
hence we can replace the sum over $n$ by an integration. Thus we may
write
\be \label{ekk7}
\wt F^{-1} \simeq 1 + \int_{-\infty}^\infty dt {m_0 \over \{\wt R^2 +
(t - x^{D-1})^2\}^{D-3\over 2}}\, ,
\ee
\be \label{ekk8}
\wt \omega_\phi \simeq \int_{-\infty}^\infty dt {m_0 a \wt s^2\over
\{\wt R^2 + (t-x^{D-1})^2\}^{D-1\over 2}}\, .
\ee
Using a change of variable $t \equiv x^{D-1} + \wt R u$ we can rewrite
the above equations as,
\be \label{ekk10}
\wt F^{-1} \simeq 1 + {C_1 m_0\over \wt R^{D-4}}\, , \qquad
\wt \omega_\phi \simeq 1 + { C_2 m_0 a \wt s^2\over \wt R^{D-2}}\, ,
\ee
where,
\be \label{ekk11}
C_1 = \int_{-\infty}^\infty {du \over (u^2+1)^{D-3\over 2}}\, ,
\qquad
C_2 = \int_{-\infty}^\infty {du \over (u^2+1)^{D-1\over 2}}\, ,
\ee
are two numerical constants. These asymptotic forms are identical to
those given in \refb{ekk4} with $D$ replaced by $(D-1)$ and with
appropriate redefinition of $m_0$ and $a$.
Thus we see that the periodic
array of the $D$ dimensional solution has the same asymptotic field
configuration as a $D-1$ dimensional rotating black hole.

This procedure
can be used to construct solutions in five and four dimensional theories
by taking periodic and doubly periodic arrays of extremal black holes in
six dimensions.
The asymptotic form of these solutions
is the same as that of rotating black hole solutions in five and
four dimensions saturating the Bogomol'nyi bound which have naked
singularities.
But now the solution near the black hole becomes six dimensional, and
the singularity is that of the extreme  rotating black holes that we
have described here.
This implies that the gyromagnetic ratios of the four dimensional solution,
which are computed from the asymptotic form of the gauge field configuration,
are identical to those of the elementary string states with the same
quantum numbers, since this equality is known to hold
for the singular solution
in four dimension\cite{GENBL}. Thus these solutions are good candidates for
describing the field configuration around elementary string states.

\sectiono{CONCLUDING REMARKS}

In this paper, we have considered only rotating black holes with one
component of angular momentum. More general solutions can be constructed
by starting with the higher dimensional Kerr solution with all components
of the angular momentum nonzero\cite{MYPE}, and
applying the solution generating
transformation.  It is likely that only some of these black holes will
saturate the Bogomol'nyi bound in their extremal limit. This needs to be
investigated further. It remains to be seen whether these black holes can be
fruitfully identified with nonspherically symmetric BPS saturated string
states. One unusual feature is that there appears to be no upper bound
on the magnitude of the angular momentum. Since $a$ is an arbitrary
parameter, the solutions we have constructed can saturate the Bogomol'nyi
bound with any value of $J$.

Although we have considered solutions to heterotic string
theory compactified on a torus, our $D=6$ rotating black hole
can easily be transformed into a Type IIA string solution. This is because
the low energy effective action for the Type IIA string theory
compactified  on K3 is related to that of the heterotic string
compactified on a torus by a simple field redefinition.
This may be useful in testing the string-string duality conjecture in
six dimensions.

\vskip 1cm

{\bf Acknowledgment}: We wish to thank J. Gauntlett and
J. Schwarz for useful discussions.
We also acknowledge the hospitality of the
Aspen Center for Physics where this work was initiated. GTH was supported
in part by NSF Grant PHY-9507065.

\end{document}